\documentclass[acmlarge]{acmart}
\usepackage{adjustbox}
\usepackage{multirow}
\AtBeginDocument{%
  \providecommand\BibTeX{{%
    \normalfont B\kern-0.5em{\scshape i\kern-0.25em b}\kern-0.8em\TeX}}}

\acmJournal{HEALTH}
\acmVolume{0}
\acmNumber{0}
\acmArticle{0}
\acmMonth{0}

\begin{document}

\title{A Review of the Non-Invasive Techniques for Monitoring Different Aspects of Sleep}




\author{Zawar Hussain}
\email{zawar.hussain@hdr.mq.edu.au}
\orcid{1234-5678-9012}
\affiliation{%
  \institution{Macquarie University}
  \streetaddress{North Ryde}
  \city{Sydney}
  \state{NSW}
  \country{Australia}
  \postcode{2109}
}

\author{Quan Z. Sheng}
\email{michael.sheng@mq.edu.au}
\orcid{1234-5678-9012}
\affiliation{%
  \institution{Macquarie University}
  \streetaddress{North Ryde}
  \city{Sydney}
  \state{NSW}
  \country{Australia}
  \postcode{2109}
}

\author{Wei Emma Zhang}
\email{wei.e.zhang@adelaide.edu.au}
\orcid{1234-5678-9012}
\affiliation{%
  \institution{The University of Adelaide}
  \city{Adelaide}
  \state{SA}
  \country{Australia}
  \postcode{2109}
}

\author{Jorge Ortiz}
\email{jorge.ortiz@rutgers.edu}
\affiliation{%
 \institution{Rutgers University}
 \city{New Brunswick}
 \state{New Jersey}
 \country{USA}}

\author{Seyedamin Pouriyeh}
\email{spouriye@kennesaw.edu}
\affiliation{%
  \institution{Kennesaw State University}
  \city{Kennesaw}
  \state{Georgia}
  \country{USA}}

\renewcommand{\shortauthors}{Zawar, et al.}

\begin{abstract}
Quality sleep is very important for a healthy life. Nowadays, many people around the world are not getting enough sleep which is having negative impacts on their lifestyles. Studies are being conducted for sleep monitoring and have now become an important tool for understanding sleep behavior. The gold standard method for sleep analysis is polysomnography (PSG) conducted in a clinical environment but this method is both expensive and complex for long-term use. With the advancements in the field of sensors and the introduction of off-the-shelf technologies, unobtrusive solutions are becoming common as alternatives for in-home sleep monitoring. Various solutions have been proposed using both wearable and non-wearable methods which are cheap and easy to use for in-home sleep monitoring. In this paper, we present a comprehensive survey of the latest research works (2015 and after) conducted in various categories of sleep monitoring including sleep stage classification, sleep posture recognition, sleep disorders detection, and vital signs monitoring. We review the latest works done using the non-invasive approach and cover both wearable and non-wearable methods. We discuss the design approaches and key attributes of the work presented and provide an extensive analysis based on 10 key factors, to give a comprehensive overview of the recent developments and trends in all four categories of sleep monitoring. We also present some publicly available datasets for different categories of sleep monitoring. In the end, we discuss several open issues and provide future research directions in the area of sleep monitoring.
\end{abstract}



\maketitle

\section{Introduction}
Sleep is a vital element in daily human life and is as essential as water and food. Humans spend one-third of their life in sleep as they need 6-8 hours of sleep daily for a healthy lifestyle.
Quality sleep is also essential. Low quality or inadequate sleep leads to impaired physical health and may cause cognitive and psychological problems. People who do not get proper sleep are more prone to different chronic diseases such as obesity, diabetes, and hypertension \cite{xie2017review}. Spiege et al. \cite{spiegel2005sleep} showed that inadequate sleep could increase the risk of obesity and diabetes. Vorona et al. \cite{vorona2005overweight} demonstrated that there is a relation between sleep time and obesity. Brooks et al. \cite{brooks1997obstructive} assert that sleep apnea is a high-risk factor for hypertension. Inadequate sleep is also one reason for road accidents and lower productivity \cite{kalsi2018sleep,ishibashi2020association}. 
Due to the importance of sleep, this area has become very popular among the research community. Many studies have been conducted to analyze and understand the quality and behaviour of sleep. It has become a separate branch of medicine and has gained a lot of attention recently \cite{wu2014assess,hussain2019cost}. 

There are various sub-topics in the area of sleep monitoring. Different studies have focused on one or more of these sub-topics. Some studies focus on sleep stages classification, and they are more interested in investigating the sleep cycles \cite{boostani2017comparative,aboalayon2016sleep}. Sleep cycles are fundamental, and any anomaly in sleep cycles or less time in a specific cycle can lead to disorders causing health problems. Some studies focus on monitoring different postures during sleep \cite{de1983sleep}. They investigate the body movements and postures such as supine, prone, left, and right and the time spent in each pose during sleep. Posture monitoring can provide useful information about sleep behaviour and is vital for preventing pressure ulcers and sleep apnea \cite{mansfield2019pressure}. Other studies focus on the detection of sleep disorders \cite{panossian2009review}. These studies investigate sleep-related disorders such as apnea, insomnia, and restless leg syndrome. The detection of these disorders is very challenging as the patients are unaware of these disorders. One other group of studies focuses on monitoring vital signs such as respiration and heart rate during sleep \cite{hussain2019cost}. Vital signs are correlated with sleep behavior and can provide important information about sleep quality and some disorders.

From the related literature, we can find that various techniques have been used to monitor sleep. The traditional approach to analyze and understand sleep quality is to use clinical methods \cite{fallmann2019computational}. Clinical methods apply specialized hardware and environment and are supervised by trained people usually, the health workers. The patient has to stay overnight in the clinic. This approach is both complex and costly.
Nowadays, smart health technologies offer ubiquitous solutions for sleep monitoring, which are relatively cheap and easy to use \cite{park2019smart,perez2020future}. Innovations in intelligent sensing have led to the development of many sensors that can be used as wearable or embedded in bed frames. These sensors can easily monitor sleep by capturing the related data. These solutions can be used for in-home sleep monitoring and do not require trained technicians.

In this paper, we review and analyze non-invasive and in-home studies conducted for monitoring sleep. We focus on techniques which do not require the users to wear different types of invasive sensors which can cause disturbance in natural sleep or require special environment. We divide the related literature into four categories: 
{\em sleep stages}, {\em sleep postures}, {\em sleep disorders}, and {\em vital signs monitoring}. We review the 
state-of-the-art works in these fields and provide the reader an overview of the advancements in these areas of sleep monitoring. We discuss and compare both wearable and non-wearable (device-free) methods used for monitoring sleep and provide a comprehensive comparison of these techniques using key factors. We also identify the possible research gaps and provide future research directions for the researchers. 

The rest of the paper is organized as follows. Section~\ref{sec:2} presents the related work, and Section~\ref{sec:3} provides the background of sleep monitoring and the clinical approaches 
used 
for sleep monitoring. We also collect and present some publicly available data sets in this section. Section~\ref{sec:4} presents the four categories of sleep monitoring in detail. We review the latest literature for both wearable and non-wearable solutions in this section. Section~\ref{sec:5} provides future research directions and Section~\ref{sec:6} offers some concluding remarks. 

\section{Related Work}\label{sec:2}

\begin{table}
\centering
\begin{adjustbox}{max width= \linewidth}
\begin{tabular}{|p {1.5 cm}|l|p {2 cm}|p {1.5 cm}|p {2 cm}|p {1.5 cm}|p {2 cm}|p {2.5 cm}|l|p {2.5 cm}|l|l|} 
\hline
\textbf{Surveys}                 &\textbf{Year} & \textbf{Main Focus} & \textbf{Data sets} & \textbf{Comparisons} & \textbf{Sleep Stages} & \textbf{Sleep Disorders} & \textbf{Vital Signs} & \textbf{Postures} & \textbf{Clinical Methods} & \multicolumn{2}{l|}{\textbf{In-Home Methods}}  \\ 
\cline{11-12}
                                                  &                                &                                      &                                     &                                             &                                        &                                           &                                       &                                    &                                            & \textbf{Wearable} & \textbf{Non-wearable}      \\ 
\hline
\cite{kelly2012recent}           & 2012                           & Consumer devices                     & No                                  & No                                          & Yes                                    & No                                        & No                                    & No                                 & No                                         & Yes               & Yes                        \\ 
\hline
\cite{surantha2016internet}      & 2016                           & Sleep quality                        & No                                  & No                                          & No                                     & Yes                                       & No                                    & No                                 & Yes                                        & No                & No                         \\ 
\hline
\cite{ibanez2018survey}          & 2018                           & Questionnaires and diaries           & No                                  & Yes                                         & No                                     & Yes                                       & No                                    & No                                 & No                                         & No                & No                         \\ 
\hline
\cite{matar2018unobtrusive}      & 2018                           & Physiological signals (actigraphy)   & No                                  & Yes                                         & Yes                                    & Yes                                       & Yes                                   & Yes                                & Yes                                        & Yes               & Yes                        \\ 
\hline
\cite{sadek2019internet}         & 2019                           & Consumer devices                     & No                                  & No                                          & No                                     & No                                        & No                                    & No                                 & No                                         & Yes               & Yes                        \\ 
\hline
\cite{fallmann2019computational} & 2019                           & Computational analysis methods       & No                                  & Yes                                         & Yes                                    & Yes                                       & No                                    & Yes                                & Yes                                        & Yes               & Yes                        \\ 
\hline
\cite{tran2019doppler}           & 2019                           & Sleep apnea                          & No                                  & No                                          & No                                     & Yes                                       & No                                    & No                                 & No                                         & No                & Yes                        \\ 
\hline
\cite{ahmadzadeh2020review}      & 2020                           & Sleep apnea                          & No                                  & Yes                                         & No                                     & Yes                                       & Yes                                   & No                                 & Yes                                        & Yes               & Yes                        \\ 
\hline
\cite{pan2020current}            & 2020                           & Wireless body area network           & No                                  & Yes                                         & Yes                                    & Yes                                       & Yes                                   & No                                 & Yes                                        & Yes               & Yes                        \\ 
\hline
Our Survey                                        & 2021                           & Ubiquitous methods                   & Yes                                 & Yes                                         & Yes                                    & Yes                                       & Yes                                   & Yes                                & Yes                                        & Yes               & Yes                        \\
\hline
\end{tabular}
\end{adjustbox}
\caption{Summary of the previous surveys and this survey}
\label{tab:table1}
\vspace{-5mm}
\end{table}

There are several surveys that summarize the works done in sleep monitoring and focus on different aspects of sleep monitoring. Kelly et al.  \cite{kelly2012recent} presented a review article summarizing the recent developments for in-home sleep-monitoring devices. This study mainly focuses on the commercially available devices and categorizes them in brain signal-based, autonomic signal-based, movement-based, and bed-based systems. This work also presents the discussion about using other consumer devices that can be used for sleep monitoring. The authors highlight research gaps in the sleep monitoring area and concluded that portable devices are increasing for sleep monitoring. Still, these devices and systems need to be standardized. Surantha et al. \cite{surantha2016internet} conducted a survey on the 
Internet of Things (IoT) related works for sleep monitoring. The emphasis of this work is on the quality of sleep. The authors discussed a few sleep-related disorders and the clinical techniques for sleep monitoring. This work presents an IoT architecture for sleep quality monitoring. Ibanez et al. \cite{ibanez2018survey} focused on sleep questionnaires and diaries. This survey only reviews the self-reporting approaches for sleep assessment and provides some critical analysis of these techniques' accuracy and validation. 

Matar et al. \cite{matar2018unobtrusive} presented a comprehensive review of the unobtrusive methods used for sleep monitoring. This survey addresses three main topics: sleep stages, sleep disorders, and monitoring techniques. The authors categorized the related literature based on physiological signals captured during sleep, such as breathing, cardiac signals, and body movements, and review the various studies in each of these categories. The authors also discussed the challenges and limitations of the techniques presented for each category.
Sadek et al. \cite{sadek2019internet} presented a survey on the use of non-intrusive technologies for sleep monitoring. The paper discusses both wearable (e.g., bracelets) and non-wearable (e.g., bed-embedded) approaches but the main focus of this work is consumer devices and various gadgets available in the market for sleep assessment. There is very less content on the research aspect of the discussed technologies. Fallmann et al. \cite{fallmann2019computational} proposed a survey on sleep monitoring techniques. This survey discusses sleep behaviour in terms of movements, stable state and disorders. The authors categorize the literature into clinical and home-based approaches and summarize different techniques from each category. However, the main focus of this survey is computational analysis methods for sleep assessment. The authors discussed various computational studies presented for sleep stages classification, body movements monitoring, and disorders detection and list the challenges and limitation of those techniques. 

Tran et al. \cite{tran2019doppler} presented a survey on the Doppler radar-based monitoring techniques for the detection of sleep apnea. This paper describes the architecture of radar-based system and discusses the different aspects of the system such as noise removal and signal processing. The authors categorized the radar-based techniques into four categories based on their methodology. These categories are time-frequency analysis which uses time-series and frequency domain in the signal processing, numerical analysis which uses numerical techniques, classification and training which uses machine learning, and other methodologies which uses experimental and mathematical modeling. Ahmadzadeh et al. \cite{ahmadzadeh2020review} 
conducted 
a survey on biomedical sensors, technologies and algorithms for the breath-related sleeping disorders. The paper categorizes the literature into invasive and non-invasive techniques and provides a review of the related articles in both the categories. The authors also discuss the different challenges involved in the sensors selection, data processing, and classification algorithm used in the presented literature. Pan et al. \cite{pan2020current} presented a review of the current techniques and future challenges in the area of sleep monitoring. The authors reviewed the related literature on sleep stages classification and sleep disorders and address some of the challenges faced by the community in the current systems for sleep monitoring. The paper also discusses clinical and wearable sensor-based approaches for sleep monitoring and proposed their own system for sleep monitoring. The proposed system consists of various wearable sensors attached to both wrists, both legs and chest position and uses machine learning-based approach to monitor sleep stages, body movements, and snoring.

There is a growing interest in the area of sleep monitoring and several surveys have been 
conducted 
recently. Table ~\ref{tab:table1} gives a summary of these surveys. As can be seen from the table, these surveys focus on different aspects of sleep monitoring like disorders detection, sleep stages classification or posture recognition during the sleep. Also, the techniques reviewed by these surveys are different and focus on one or more aspects of sleep monitoring. None of these surveys discussed the datasets for sleep monitoring which is a very important part of research in this field. To the best of our knowledge, there is no survey paper which provides a complete review of the work conducted in all the four sub-categories of sleep monitoring using the non-invasive approach. Due to this reason, we 
believe 
that 
there is a need for a comprehensive review to give readers an overview of the latest 
research efforts 
in all the sub-categories of sleep monitoring, especially using the unobtrusive techniques. In this paper, we review the recent works in sleep monitoring in the four sub-categories of sleep monitoring which are sleep stages classification, sleep disorders detection, posture recognition and vital signs monitoring.

\section{Background}\label{sec:3}
In this section, we provide the background of sleep monitoring. We also discuss the clinical methods used for sleep monitoring.
Due to the well-recognized importance of quality sleep and high prevalence of inadequate sleep, sleep monitoring has become a very important area of research. Many studies have been conducted which proposed different techniques for monitoring different aspects of sleep. 

\begin{figure}[tb!]
	\centering
	\includegraphics[width= \linewidth]{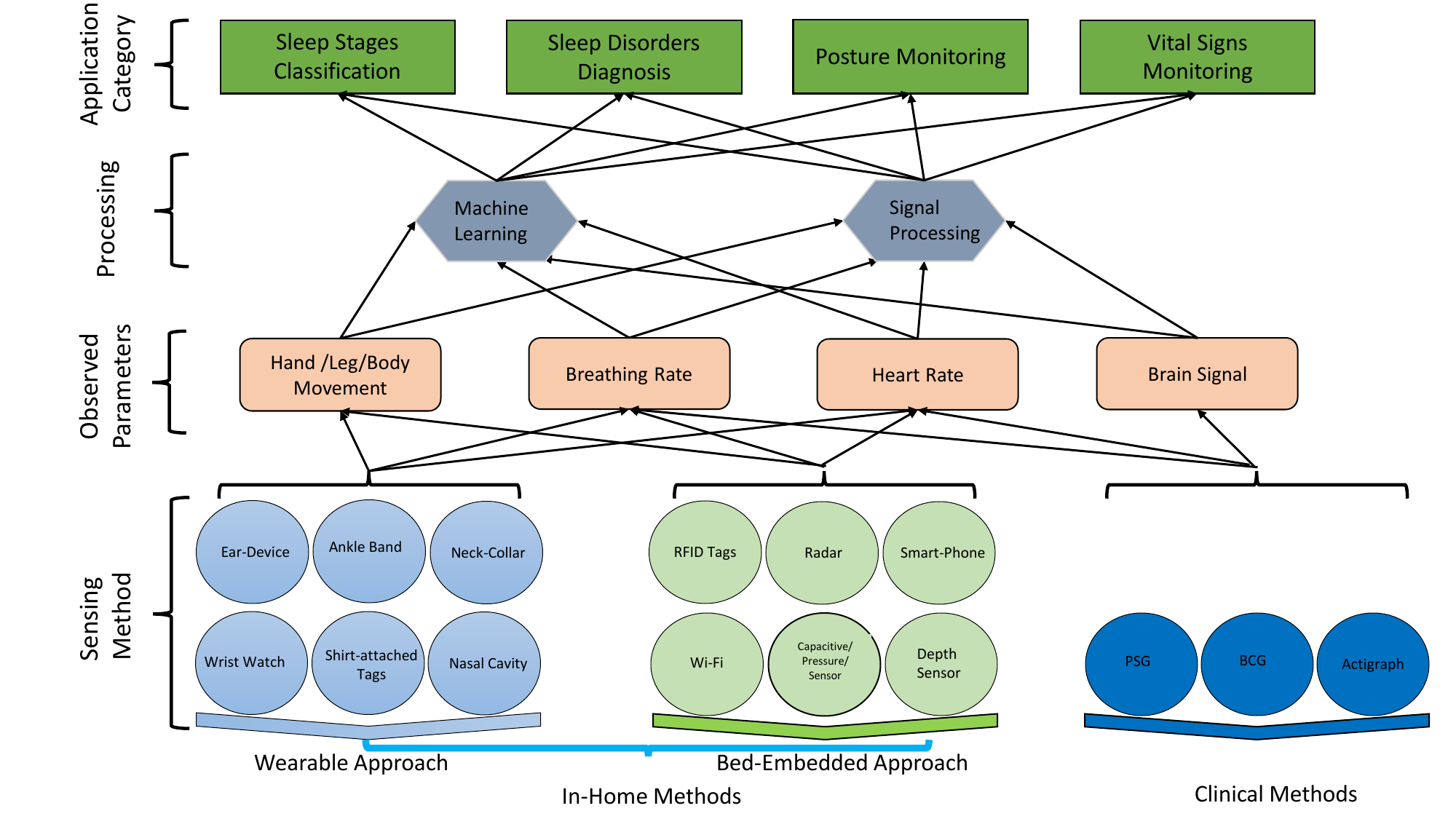}
	\caption{General architecture of sleep monitoring system.}
	\label{fig:figure1}
\end{figure}

The general architecture of a sleep monitoring system can be divided into four layers as shown in Figure ~\ref{fig:figure1}, 
consisting 
1) sensing, 2) observed parameters, 3) processing, and 4) application. 
The first layer deals with the choice of sensing method. Different sensors have been used in the literature to capture the information about the sleeping. In clinical methods, various sensors are used in an invasive method to capture the physiological signal while the subject is asleep. In-home methods use both wearable and bed-embedded devices. In the wearable method, different sensors are worn by the subjects on their body at different positions such as wrist, leg, ankle, chest and neck. The sensors (e.g., pressure, capacitive, RFID tags)
can be embedded in the bed frame or the sleeping mattress as well.
The second layer of the architecture deals with the type of parameters observed. The sleeping involves various factors such as body movements, hand movements, leg movements, breath rate, heart rate and brain signals. Based on the objective of the system, one or more of these parameters can be captured. The third layer consists of the processing of the data captured. Two main approaches are used in the literature for processing the data. These approaches are machine learning and signal processing. The final layer consists of application category in which the final output of the system is decided. The output can be one or more of the given four categories.

\subsection{Clinical Approaches for Sleep Monitoring}
Clinical methods refer to those methods which are conducted in special environments (labs/clinics/hospitals) under the supervision of trained staff. Currently, these methods have the highest accuracy and are considered as standards for sleep monitoring. These methods can be termed as invasive because they involve various sensors attached to different body parts of the subject (e.g., head, nose, fingers, toes, and chest).
These sensors cause disturbance in the natural sleep and cannot be used for long term monitoring. In this section, we discuss some of the important clinical methods used for sleep monitoring.

\subsubsection{Polysomnography}
\begin{figure}[!tb]
	\centering
	\includegraphics[width= 3 in]{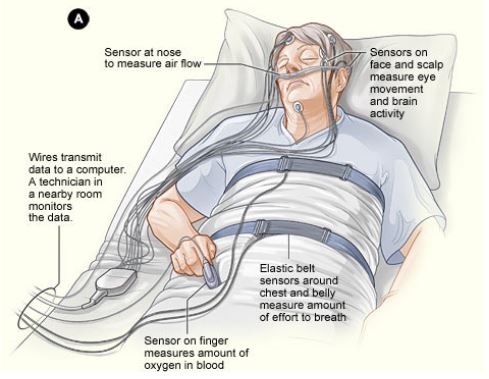}
	\caption[]{Polysomnography\footnotemark}
	\label{fig:figure2}
\vspace{-3mm}
\end{figure}

The current gold standard technique used for sleep analysis and diagnosis of sleep disorders is Polysomnography (PSG) \cite{sadek2017nonintrusive}. PSG is a complex clinical process which requires the subject to sleep for hours wearing many sensors as shown in Figure ~\ref{fig:figure2}. 
 This procedure can be conducted in specialized environment under the supervision of trained people (sleep experts and technicians). The patient has to stay overnight and the data from electroencephalogram (EEG), electrocardiography (EOG), electromyography (EMG), nasal airflow, and pulse oximetry is recorded via the attached sensors. The data is then processed to analyze the sleep behaviour. PSG is not an ideal procedure for sleep analysis as the subject has to wear many different kinds of sensors and has to stay overnight in the specialized environment. This procedure is time-consuming, complex, and expensive and cannot be used for long-term \cite{fang2018novel}. 

\footnotetext{Source: https://www.sleep-apnea-guide.com/polysomnogram.html}

\subsubsection{Actigraphy}

Actigraphy is another approach which has gained popularity in health care solutions \cite{smith2018use}. Actigraphy is widely accepted as standard in medical domain for objective sleep quality measurement \cite{fallmann2019computational}. In this approach, a device which mainly consists of an accelerometer (can also use other inertial sensors such as gyrsocope or magnetometer) is attached to the body (preferably wrist or ankle position) of the person which can record all the physical movements of the person. The normal assessment consists of seven consecutive days to get a representative image of the user's sleep. The data is recorded and then processed using machine learning or signal processing techniques for sleep analysis. This method has the advantage that it can monitor the person's sleep for longer period without any extra cost or effort. However,
it can only provide information about the wake and sleep activity labels and the physical movements during the sleep \cite{schwab2018actigraphy}.

\subsubsection{Ballistocardiography}
Ballistocardiography (BCG) is another clinical technique which is based on the signal generated by the heart when it pushes the blood into the vessels. This is a very old technique but recently got significant importance and has been used for vital signs and sleep monitoring \cite{giovangrandi2011ballistocardiography}. This approach provides non-invasive monitoring of cardiac signals and does not require the users to wear any sensors on their bodies. Different sensors are employed to capture BCG signals such as fiber brag grating sensors, electromechanical film sensors, piezoelectric film sensors, microbend fiber optic sensors, polyvinylidene-fluoride sensors, hydraulic and pneumatic sensors, and pressure sensors \cite{inan2015ballistocardiography}\cite{kim2016ballistocardiogram}. These sensors can be embedded in ambient environment such as bed and chairs. BCG method is mostly used for monitoring the cardiac activity  and can provide good results but it involves a lot of sensors to be deployed which makes it costly. Also, these sensors can be affected by different factors such as bed frames, mattress thickness, body movements, bed-partners and motion artifacts.

\subsection{Publicly Available Data Sets for Sleep Monitoring}
Nowadays, there are many data sets publicly available in various research areas. These data sets are playing a vital role in the advancement of research as the  research community can use these data sets to evaluate their newly proposed solutions. Like other fields of science, there are publicly available data sets for sleep monitoring as well. Most of these data sets consist of PSG recordings and are collected in specialized environment (hospitals/clinics). The National Sleep Research Resource (NSRR)\footnotemark offers free access to many sleep related data-sets. Table ~\ref{tab:table2} provides details about some of the publicly available data sets for sleep monitoring.

\footnotetext{Source: https://sleepdata.org/datasets}

\begin{table}[!bt]
\centering
\begin{adjustbox}{max width= \linewidth}
\begin{tabular}{|p {1 cm}|l|p {2.5 cm}|p {4 cm}|p {3 cm}|p {1.5 cm}|p {3 cm}|p {1 cm}|} 
\hline
\textbf{Data set}                                & \textbf{Year} & \textbf{Focus}      & \textbf{Sensor Used}                                         & \textbf{Deployment}        & \textbf{\# Subjects} & \textbf{Sample Size}                 & \textbf{Age range}  \\ 
\hline
\cite{quan1997sleep}            & 1997          & OSA,Sleep stages    & PSG                                                          & Specialized environment    & 6000                 & NA                                   & 40                  \\ 
\hline
\cite{goldberger2000physiobank} & 2000          & General purpose     & PSG                                                          & Specialized clinic         & 16                   & 8-10 hours per subject               & 32-56               \\ 
\hline
\cite{penzel2000apnea}          & 2000          & Apnea               & ECG                                                          & Specialized clinic         & 32                   & 34313 minutes                        & NA                  \\ 
\hline
\cite{heneghan2011st}           & 2011          & Apnea, sleep stages & PSG (Jaeger-Toennies system)                                 & Specialized clinic         & 25                   & 5.9
to 7.7 hours per subject         & 28-68               \\ 
\hline
\cite{o2014montreal}            & 2014          & General purpose     & PSG                                                          & Specialized clinic         & 200                  & One full night per subject           & 18-76               \\ 
\hline
\cite{khalighi2016isruc}        & 2016          & General purpose     & PSG                                                          & Specialized clinic         & 100 + 8 + 8          & 1, 2, and 1 session for each subject & NA                  \\ 
\hline
\cite{pouyan2017pressure}       & 2017          & Sleep postures      & Pressure sensor (Vista Medical FSA SoftFlex 2048)            & Sleeping mat under mattress & 13                   & 23400                                & 19-34               \\ 
\hline
\cite{zhao2017learning}         & 2017          & Sleep stages        & Radio                                                        & Room                       & 255                  & 90000 (30-sec epoch)                 & NA                  \\ 
\hline
\cite{8743916}                  & 2018          & Arousal             & PSG                                                          & Specialized clinic         & 994                  & 1983                                 & 55 (mean)           \\ 
\hline
\cite{hillyard2018experience}   & 2018          & Respiration         & RF (UWB-IR,WiFi CSI, Zigbee RSS, and sub-1 dB quantized RSS) & Either sides of the bed   & 20                   & 160 hours                            & 55-60               \\
\hline
\end{tabular}
\end{adjustbox}
\caption{Summary of the publicly available data sets}
\label{tab:table2}
\vspace{-5mm}
\end{table}

\section{Sleep Monitoring} \label{sec:4}
There are four categories of sleep monitoring (see Figure~\ref{fig:figure1}). 
In this section, we review the latest work conducted in these categories.

\subsection{Sleep Stages Monitoring}

A very important aspect in sleep monitoring is the classification of sleep stages which can help in the analysis of the sleep behaviour and diagnosis of sleep disorders. Initially, the sleep was divided into six stages by the R-K method proposed by Rechtschaffen and Kales in 1968 \cite{rechtschaffen1968manual}. Later in 2007, the American Academy of Sleep Medicine (AASM) \cite{iber2007aasm} updated their manual for sleep scoring and divided the sleep into five stages. These stages are wake, rapid eye movement (REM) and three stages (N1, N2, and N3) of non-rapid eye movement (NREM) as shown in Table~\ref{tab:table3}. Sleep cycle in healthy adults last for around 90 to 100 minutes and it starts with N1 stage which is the shortest followed by N2 (which lasts for around 10 to 25 minutes) and 
then N3 stage. 
After N3, 
the REM stage 
counts for 20-25 \% of the total sleep. Monitoring and identification of these sleep stages is very critical for diagnosis of different sleep disorders and it can also help in tracking the treatment's response in patients \cite{zhao2017learning}. In this section, we review the work conducted for sleep stages classification using the ubiquitous approach.

\begin{table}[!h]
\centering
\small
\begin{adjustbox}{max width= \linewidth}
\begin{tabular}{|p {2 cm}|p { 1 cm}|p { 10 cm}|} 
\hline
\multicolumn{2}{|c|}{\textbf{Stage}} & \textbf{Description}                                                                                                                                                                   \\ 
\hline
Wake                     & W         & The person is awake but on the bed                                                                                                                                                     \\ 
\hline
\multirow{3}{*}{NREM}    & N1        & Dosing off which lasts for 1 to 5 minutes                                                                                                                                              \\ 
\cline{2-3}
                         & N2        & A more subdued state in which the body is relaxed. The body temperature drops and the  breathing, heart rate, and brain activity starts to slow down. N2 can last for 10-25 minutes    \\ 
\cline{2-3}
                         & N3        & Deep sleep. Muscles are relaxed and breathing and brain activity slow down. N3 can last for 20-40 minutes and is the most important part of the sleep for different bodily processes.  \\ 
\hline
Rapid Eye Movement (REM) & REM       & In this stage, the whole body is paralyzed except the eyes and the brain. Dreams occur at this stage. REM is responsible for around 25\% of the total sleep in adults.                 \\
\hline
\end{tabular}
\end{adjustbox}
\caption{Five stages of sleep}
\label{tab:table3}
\vspace{-3mm}
\end{table}

\subsubsection{Wearable-based Approaches for Sleep Stage Classification}
Due to the advancements in the production of wearable devices, the use of wearables has been increased significantly. Many applications especially, human activity recognition \cite{lu2019mfe,hussain2020review} and healthcare \cite{adame2018cuidats} are now adopting the wearable approach because of the low-cost and non-complex operation. In this section, we review the wearable-based techniques used for sleep stage classification.

Beattie et al. \cite{beattie2017estimation} conducted a study for sleep stage estimation using the optical pulse plethysmography (PPG) and accelerometer. They collected the overnight sleep data from 60 adults who wore the devices (Fitbit) on both of their wrists. The authors use the Embletta for validation (ground truth) purposes which collects the EEG, ECG and EOG signals. Two sleep experts were hired to score the PSG records from the Embletta device as per the AASM guidelines. The labeling from the experts is used as the ground-truth for the valuation of the proposed approach. This work uses three sets of features which are from movement, heart rate and respiration. These feature are fed into a linear discriminant classifier which can predict one of the four classes (Wake, REM, Light, and Deep) with 69\% accuracy. One of the main findings of this study is that actigrapghy can provide some information about the sleep stages (such as awake and sleep) but cannot accurately predict all the stages. By adding the physiological data (heart rate and breath rate) captured by PPG, the sleep stages can be detected with reasonable accuracy. Zhang et al. \cite{zhang2018sleep} proposed a techniques for sleep stage classification based on actigraphy and heart rate captured by a wearable device. To get the ground truth, five sleep experts analyzed the PSG data and assigned the five sleep stages independently. The low-level features in both time and frequency domain are extracted from accelerometer and heart rate data which are then combined to learn the mid-level features. A bidirectional long short-term memory (BLSTM) architecture is trained on the final features to classify the sleep data into one of the five stages. The authors evaluate the performance of the proposed system for two different groups; the resting group went to sleep without doing any exercise while the comprehensive group exercised before going to bed. The achieved an F1 score of 64\%for the resting group and 60.5\% for the comprehensive. 

Boe et al. \cite{boe2019automating} presented a technique based on wearable wireless senors to detect different stages of the sleep. This work uses IMU devices attached to different positions on the body and collects the physiological data including the body temperature during the sleep. PSG is used to get the ground truth. A population-based bagging classifier with decision tree (DT) is used to classify the sleep data into different classes. This work has evaluated the proposed approach for two-stage (wake vs sleep), thee-stage (wake vs NREM vs REM) and four-stages classification (wake vs light vs deep vs REM) and the performance was compared with that of Actiwatch. The results show that their approach outperforms the Actiwatch but the performance of the proposed technique is poor for the four-stage classification. Zhang et al. \cite{zhang2019sleep} designed their own wearable multi-sensor vest called SensEcho for sleep stage classification. The SensEcho uses two electrodes attached to the chest and abdomen region and one oximeter attached at wrist position. An accelerometer is also embedded in the vest. This multi-sensor vest captures the ECG and the breathing signal. A number of different features including both time and frequency domain features are extracted form the raw data. The extracted features are fed into a BLSTM which can assign one of the four stages with 80.75\% accuracy. The authors also evaluated the proposed approach on a publicly available data set and achieved the same accuracy. Nakamura et al. \cite{nakamura2020hearables} proposed another wearable technique called Hearables which uses an in-ear sensor for sleep stages classification. This in-ear device has two electrodes and can easily record the ECG signal for the over-night sleep. After the pre-processing, different features are extracted which includes structural complexity features and spectral features. The final features are fed into an SVM with radial basis function (RBF) kernel which classify the sleeping data into five classes with 74\% acuracy. PSG is used as a ground truth.

\subsubsection{Non-Wearable-based Approaches for Sleep Stage Classification}
A major issue with the wearable approach is that users are bound to wear the device all the time. Nowadays, many solutions are using the device-free (non-wearable) approach \cite{hussain2020review} in which the users are not required to wear anything. Non-wearable devices-based approaches have the advantage of having the least disturbance in natural sleeping because the users are not required to wear any sensor. The sensors devices are either placed in the environment (room) or embedded in the bed frame or mattress. Sleep monitoring can be done via a single device such as radar or multiple sensors can be integrated in the bed. The most commonly used sensors which are  used as non-wearable for sleep monitoring include pressure sensor, radar, Wi-Fi, radio, load cells, force sensors, smart phones etc. In this section, we discuss the non-wearable-based techniques used for sleep stage classifications.

Zhao et al. \cite{zhao2017learning} proposed a radio signal-based technique for predicting the sleep stages without attaching anything to the subject body. This technique uses a radio device placed in the bedroom while the user is sleeping and then analyzes the reflected signals to extract the information about the sleep. One of the main contribution of this work is the feature engineering. The authors use game theory and adversarial training for discarding all the irrelevant information in the reflected signal and extract only the relevant features. This work uses a combination of Convolutional Neural Network-Regular Neural Network (CNN-RNN) in their model architecture and achieves an accuracy of 79.80\%. Zhang et al. \cite{zhang2017sleep} presented a non-contact technique for sleep stage classification using the Doppler radar. A continuous-wave Doppler radar is placed near the user bed which emits a single tone signal. The reflected signal contains the information about the breathing of the user because the chest movements induce disturbances in the signal. The reflected signal also captures the movements of the user. Different features related to body movements, respiration rate and heart beat are extracted from the reflected signal which are fed into a bagged tree classifier. The classifier can assign one of the four classes with 78.6\% accuracy. 

Yi et al. \cite{yi2019non} proposed the use of a hydraulic sensors for monitoring the sleep stages in a non-invasive way. The hydraulic sensors are embedded in the bed under the mattress which can capture the body movements as well as
 the movements caused by breathing and BCG. 
 The captured data is first cleaned by passing through Butterworth filter to remove the low and high frequency noise. Different features including both time and frequency domain are calculated for respiration and heart rate. Two classifiers, Support Vector Machine (SVM) and k-nearest neighbors (KNN), are used in both single layer and hierarchical configurations to predict the three stages of sleep. Single layer SVM achieved the highest accuracy of 85.5\%. A similar approach is presented by Gargees et al. \cite{gargees2019non} which also uses the hydraulic sensors embedded in the bed frame to classify the different sleep stages. After cleaning the BCG data captured by the hydraulic sensors, it is passed through a deep model architecture which consists of CNNs and LSTMs. A pre-trained CCN (trained on the sleep data of 56 subjects from another study) is employed using the transfer learning technique. The study shows that transfer learning can boost the performance of the classifier and can achieve an accuracy of 90.80\%. Zhang et al. \cite{8826242} presented the design and implementation of a system called SMARS which uses ambient radio signals for sleep stage classification. The basic working principle of this system is the tiny changes induced in the reflected signal due to breathing and body movements. This work employs a statistical model which accounts for all the reflecting and scattering multi-paths to accurately estimate the breath rate and hence the sleep stages. Awake and sleep stages are detected based on the frequency of movement i.e., frequent movements means the person is awake. For the classification of REM and NREM stages, an SVM with RBF kernel is used 
 which 
 can predict the two classes with 88\% accuracy. 

\begin{table}
\centering
\begin{adjustbox}{max width= \linewidth}
\begin{tabular}{|p{1.5 cm}|p{1.2 cm}|p{3 cm}|p{2.5cm}|p{3 cm}|l|l|p{1.5 cm}|p{2 cm}|p{1.3 cm}|p{2 cm}|} 
\hline
\textbf{Work}                                 & \textbf{\# Stages} & \textbf{Sensor-Placement}               & \textbf{Device Used}                                               & \textbf{Methods Used}          & \textbf{Accuracy} & \textbf{Obtrusiveness} & \textbf{Cost} & \textbf{Parameter Observed}     & \textbf{\# Subjects} & \textbf{Validation Method}  \\ 
\hline
\cite{beattie2017estimation} 2017 & 4                     & Accerelerometer, PPG-wrist                      & Fitbit                                                             & Linear discriminant classifier & 69\%              & Low                    & Low           & Movement, breathing, heart rate & 60                      & PSG by Embletta MPR         \\ 
\hline
\cite{zhang2018sleep} 2018        & 5                     & Accelerometer, PPG- wrist                        & Microsoft Band I                                                   & BLSTM-based RNN                & 60.50\%           & Low                    & Low           & Movement, heart rate            & 39                      & PSG by Compumedics.    \\ 
\hline
\cite{boe2019automating} 2019     & 2,3,4                 & IMU, Temperature- chest, wrist, ankles and forehead & BioStampRC, Thermochron iButton, ActiWatch Spectrum (for comparison) & Bagging classifier with DT     & 60.60\%           & Medium                 & Medium        & Movement, temprature,ECG        & 11                      & PSG                         \\ 
\hline
\cite{zhang2019sleep} 2019        & 4                     & Accerelerometer, ECG-chest, oximeter-wrist       & SenseEcho                                                          & BLSTM-based RNN                & 80.75\%           & Medium                 & Medium        & Breathing, ECG                  & 32                      & PSG                         \\ 
\hline
\cite{nakamura2020hearables} 2020 & 5                     & EEG-ear                                         & Self-made                                                          & SVM  with RBF kernel           & 74.10\%           & Low                    & Medium        & ECG                             & 22                      & PSG by SOMNO screen         \\
\hline
\end{tabular}
\end{adjustbox}
\caption{Comparison of the wearable-based solutions for sleep stage classification}
\label{tab:table4}
\vspace{-3mm}
\end{table}

\subsubsection*{\textbf{Summary}}
A summary of the work presented for sleep stage classification is given in Table~\ref{tab:table4} (wearable-based) and 
Table~\ref{tab:table5} (non-wearable-based). From the presented work, we can see that solutions with higher level of obtrusiveness and cost have the better accuracy. Also, these systems can recognize more stages than solutions with low level of obtrusiveness. This can be explained by the level of physiological signal captured by these complex systems. Another observation is the increasing use of machine learning and feature engineering in this area. More solutions are now focusing on the feature engineering aspect and then use the existing models. Accelerometer is the widely used sensor in wearable solutions while non-wearable solutions are mostly using the pressure sensors. Although, the wearable systems are easy to use, the accuracy is low. More work is required to increase the accuracy of these solutions to an acceptable level.

\begin{table}[!tb]
\centering
\begin{adjustbox}{max width= \linewidth}
\begin{tabular}{|p{1.5 cm}|p{1.2 cm}|p{3 cm}|p{2.5cm}|p{3 cm}|l|l|p{1.5 cm}|p{2 cm}|p{1.3 cm}|p{2 cm}|} 
\hline
\textbf{Work}                                 & \textbf{\# Stages} & \textbf{Sensor-Placement} & \textbf{Device Used}     & \textbf{Methods Used}                       & \textbf{Accuracy} & \textbf{Obtrusiveness} & \textbf{Cost} & \textbf{Parameter Observed}                            & \textbf{\# Subjects} & \textbf{Validation Method}  \\ 
\hline
\cite{zhao2017learning} 2017 & 4                  & Radio-room                & COTS                     & Game theory, CCN, RNN, adversarial training & 79.80\%           & Low                    & Low           & Breathing, heartbeat from reflected signal             & 25                   & EEG-based sleep monitor     \\ 
\hline
\cite{zhang2017sleep} 2017   & 4                  & Radar-room                & Digital-IF Doppler radar & Bagged tree                                 & 78.60\%           & Low                    & Low           & Movement, respiration, heartbeat from reflected signal & 1                    & PSG                         \\ 
\hline
\cite{yi2019non} 2019        & 3                  & Pressure sensor-bed       & Hydraulic bed system     & SVM with cubic kernel, KNN                  & 85\%              & Low                    & Medium        & Respiration, heartbeat                                 & 5                    & PSG                         \\ 
\hline
\cite{gargees2019non} 2019   & 3                  & BCG, bed                  & Hydraulic bed system     & Transfer learning, CNN, LSTM, DNN           & 90.80\%           & Low                    & Medium        & Respiration, heart rate                                & 5                    & PSG                         \\ 
\hline
\cite{8826242} 2021          & 3                  & Radio-room                & Atheros WiFi chipsets.   & SVM with RBF, Statistical model             & 88\%              & Low                    & Low           & Movement, respiration from refelcted signal            & 6                    & PSG, EMFIT, ResMed          \\
\hline
\end{tabular}
\end{adjustbox}
\caption{Comparison of the non-wearable-based solutions for sleep stage classification}
\label{tab:table5}
\end{table}

\subsection{Sleep Posture Monitoring}
Recognizing sleep 
postures
is a very important factor in the assessment of sleep. It provides very useful information about the quality of sleep. Posture recognition is of particular importance for bedridden patients to avoid pressure ulcers also called pressure sores. Prolonged stay for elder people and post-surgery patients in the bed can cause skin problems due to laying in one posture for longer time. Monitoring the postures can help the care givers to provide better support and avoid the pressure ulcers and other skin problems. Due to its importance, many studies have been conducted to monitor the sleeping postures. In this section, we review some of the recent works done for posture recognition.

\subsubsection{Wearable-Based Approaches for Posture Monitoring}
In this section, we discuss the wearable-based techniques used for sleep posture monitoring.
Sun et al. \cite{sun2017sleepmonitor} presented a sleep monitoring system called SleepMonitor by using the accelerometer embedded in a smartwatch. The working of this technique is based on the idea that wrist position is correlated with the body postures. SLeepMonitor uses the wrist positions to find the sleep postures. Different statistical features are extracted from the accelerometer data which are used to train four different classifiers including NB, BN,DT, and RF. The RF performs better than the other models and recognizes the four sleep postures with 91.70\% accuracy. Fallmann et al. \cite{fallmann2017wearable} also used the accelerometer-based approach for sleep posture recognition but this work attaches the senors to both ankles and the chest. This work uses Matrix Learning Vector Quantization (GMLVQ) model, which is a distance-based classifier, to recognize one of the eight classes. 

SleepGuard \cite{chang2018sleepguard} also uses an IMU embedded in the smartwatch to monitor different sleep related activities including the postures. SleepGuard uses the arm position to recognize different sleep postures. In addition to recognize the four basic sleep postures, SleepGuard can also recognize three hand positions. Moreover, the proposed system can monitor other important sleep factors such as body rollover, micro movements, acoustic events (snoring, coughing etc.), and illumination condition. The work conducted by Jeon et al. \cite{jeon2019wearablposition} uses wearable devices attached to both wrists and the chest. The wearable device consists of IMU sensors. A two-level classifier based on Dynamic State Transition (DST)-framework is used to classify the sleep postures and motions. First, the data is classified into Sleep Posture Change (SPC) and NonSPC groups using Random Under Sampling Boost (RUSBoost)
algorithm. In the second level, SVM and RF algorithms are used to further classify SPC into eight and NonSPC into four motions. iSleePost \cite{jeng2021wrist} is another non-invasive system which uses inertial sensors attached to the wrist and chest positions of the users. The chest-attached sensor is only used in the training phase and is no longer needed after the model is trained. Basic features are extracted form the accelerometer data and two machine learning models (SVN and RF) are used to detect the four sleeping postures.

\subsubsection{Non-Wearable-Based Approaches for Posture Monitoring}
In this section, we present the non-wearable-based techniques used for sleep posture monitoring.
Xu et al. \cite{xu2016body} presented a matching-based approach 
called Body-Earth Mover's Distance (BEMD) for sleep posture recognition using the pressure image captured by the pressure sensors embedded in the bed sheet. BEMD treats the pressure image as 2D weighted image and calculates three descriptors (planar, polar and projection) before applying the histogram normalization. Euclidean distance and EMD are combined to evaluate the similarity of sleep postures. Finally, a skew-based posture classifier is used to classify the postures in one of the six postures. The skew-based classifier consists of KNN and skew rate classifier. Barsocchi et al. \cite{barsocchi2016unobtrusive} presented an unobtrusive system for posture recognition using the force sensing resistors (FSR) embedded in the bed frame. FSR captures the pressure caused by the human body (when lying in the bed). A multi-class logistic regression model is trained to classify the pressure data into one of the four postures. 

Hu et al. \cite{hu2018non} proposed an RFID-based system to recognize sleep postures. This work attaches multiple (8x8) passive RFID tags to the blanket which can capture the posture information of the sleeping person. When the person moves or change sides in the bed, the RSSI of the received signal at the reader will be disturb accordingly. This work exploits the changes caused in the received signal as a result of the posture changes and apply CNN after removing the noise. The CNN can divide RSSI image (each pixel corresponds to a tag value) into one of the five postures with 86\% accuracy. Liu et al. \cite{liu2019tagsheet} proposed a similar approach called TagSheet in which they embed multiple passive RFID tags into the sleeping mat/bed sheet in the form of a matrix. The system takes grey-scale snapshots of the variance in the RF signal from all the tags which can represent the different postures. This work employs various image processing techniques such as Gaussian blur, Ostu-based binary conversion of the grey-scale image and the removal of scattered pixels. TagSheet then applies multi-level hierarchical recognition by first classifying the postures into three coarse-grained classes and then refine it by further classifying to get the total six classes. One of the advantage of this system is that no prior data or training is required and the system can be used in a plug and play manner. 

Yue et al. \cite{yue2020bodycompass} conducted a study exploring the wireless signals and proposed a system called BodyCompass for sleep posture monitoring in a home environment. BodyCompass uses an FMCW radio which transmits a low power signal. These signals are reflected from the environment as well as the person body. BodyCompass analyzes these reflected signals by using different signal processing and machine learning techniques to infer the sleep postures. This work calculates the sleep postures in terms of angle, i.e., the angle between the bed surface and the trunk of the sleeping person. If the angle is 0, it means the posture is facing upward while 90 angle means posture is facing rightwards. BodyCompass first extracts filtered multi-path features specific to a person from the RF reflection. A fully connected neural network is trained on the data of a single user in a specific home. Then transfer learning is used to train the model for predicting the posture of new users. The use of transfer learning reduces the requirement of training data for new users. BodyCompass is evaluated in various settings and achieves an accuracy of around 94\%.

\begin{table}
\centering
\begin{adjustbox}{max width= \linewidth}
\begin{tabular}{|p{1.5 cm}|p{1.5 cm}|p{2.5 cm}|p{2.5cm}|p{3 cm}|l|l|p{1 cm}|p{2 cm}|p{1.3 cm}|p{2.5 cm}|} 
\hline
\textbf{Work}                                     & \textbf{\# Posture} & \textbf{Sensor-Placement}                                    & \textbf{Device Used}             & \textbf{Methods Used}                                                                & \textbf{Accuracy} & \textbf{Obtrusiveness} & \textbf{Cost} & \textbf{Parameter Observed} & \textbf{\# Subjects} & \textbf{Validation Method}                              \\ 
\hline
\cite{sun2017sleepmonitor} 2017  & 4                   & Accelerometer- wrist                                         & Sony Smartwatch 3, Huawei Watch & NB, BN, DT, RF                                                                         & 91.70\%           & Low                    & Low           & Wrist position              & 16                   & Smartwatch attached at belly, Comparison with BioWatch  \\ 
\hline
\cite{fallmann2017wearable} 2017 & 8                   & Accelerometer-legs, chest                                    & Shimmer3                         & Generalized Matrix Learning Vector Quantization                                      & 98\%              & Medium                 & Low           & Body position               & 6                    & Video camera                                            \\ 
\hline
\cite{chang2018sleepguard} 2018  & 4+3                 & Accelerometer, gyroscope, orientation sensor-wrist           & Self-made                        & Euclidean distance, template matching                                                & 90\%              & Low                    & Low           & Arm position                & 15                   & Video camera                                            \\ 
\hline
\cite{jeon2019wearable} 2019     & 4+8                 & Accelerometer, gyroscope, magnetometer-both wrists and chest & AHRS EBIMU24GV                   & Directional features, two-level classification, Random Under Sampling Boost, SVM, RF & 95\%              & Medium                 & Low           & Body position               & 5+11                 & Kinect                                                  \\ 
\hline
\cite{jeng2021wrist} 2021        & 4                   & Accelerometer- wrist, chest                                  & Koala                            & SVM, RF                                                                              & 85\%              & Medium                 & Low           & Body position               & 1                    & Accelerometer                         \\ 
\hline
                                                                                        
\hline
\end{tabular}
\end{adjustbox}
\caption{Comparison of the wearable-based solutions for sleep posture recognition}
\label{tab:table6}
\end{table}

\begin{table}
\centering
\begin{adjustbox}{max width= \linewidth}
\begin{tabular}{|p{1.5 cm}|p{1.5 cm}|p{2.5 cm}|p{2.5cm}|p{3 cm}|l|l|p{1 cm}|p{2 cm}|p{1.3 cm}|p{2.5 cm}|} 
\hline
\textbf{Work}                                         & \textbf{\# Posture} & \textbf{Sensor-Placement}   & \textbf{Device Used}              & \textbf{Methods Used}                                                  & \textbf{Accuracy} & \textbf{Obtrusiveness} & \textbf{Cost} & \textbf{Parameter Observed}         & \textbf{\# Subjects} & \textbf{Validation Method}  \\ 
\hline
\cite{xu2016body} 2016               & 6                   & Pressure sensor-bedsheet    & Piezo-electrical pressure sensors & EMD, Euclidian distance, KNN, Skew rate classifier                     & 91.21\%           & Low                    & Medium        & Pressure image                      & 14                   & NA                          \\ 
\hline
\cite{barsocchi2016unobtrusive} 2016 & 4                   & Force sesning resistors-bed & Self-made                         & Logistic regression                                                    & 91.20\%           & Low                    & Medium        & Pressure values                     & 3                    & Video                       \\ 
\hline
\cite{hu2018non} 2018                & 4                   & RFID tags-blanket           & COTS                              & CNN                                                                    & 86.24\%           & Low                    & Low           & Breathing from the reflected signal & 1                    & NA                          \\ 
\hline
\cite{liu2019tagsheet} 2019          & 6                   & RFID tags-bedsheet          & Impinj H47                        & Image processing, polinomial fitting, hierarchical classification, PCA & 96.70\%           & Low                    & Low           & Breathing from the reflected signal & 12                   & Video                       \\ 
\hline
\cite{yue2020bodycompass} 2020       & NA                  & Radio-room                  & FMCW radio                        & Filtered multipath extraction, majority voting, fully  Connected NN    & 94\%              & Low                    & Low           & Breathing from the reflected signal & 26                   & Accerelemoter               \\
\hline
\end{tabular}
\end{adjustbox}
\caption{Comparison of the non-wearable-based solutions for sleep posture recognition}
\label{tab:table7}
\end{table}

\subsubsection*{\textbf{Summary}}
Table~\ref{tab:table6} and Table~\ref{tab:table7} provide a summary of the work presented for sleep postures monitoring. Like other categories of sleep monitoring, the wearable-based posture monitoring solutions are mostly using the accelerometer while the non-wearable based solutions are using pressure sensor and RF technology. Different solutions are able to recognize different postures ranging 
from 
4 to 12 postures but almost all solutions presented can recognize the four basic postures with high accuracy. Validation is one concern for posture recognition system as there is no benchmark data set available and most solutions either have to use a video camera to record the ground truth or use another device to get the ground truth. Video cameras have the privacy concern while other devices may not give the 100\% accurate results. Machine learning plays an important role in posture recognition as well. Data can be collected through various sensors but it is the machine learning which can recognize the postures. Some of the commonly used machine learning algorithms in posture recognition are KNN, Random Forest, SVM, Naive Bayes (NB), and HMM.

\subsection{Vital Signs Monitoring}
Monitoring vital signs such as respiration rate (RR) and heart rate (HR) during sleep is very important. These signs provide information about the overall health of the person and can be used as indicators for the diagnosis of  many  disorders. RR can help in the diagnosis of different sleep disorders. Different  people  have  different  HR, 
depending on person’s health and age. 
An abnormal  HR may  indicate  the  possibility  of  some  kind  of health  issues such as heart problem.  According to WHO, around 17 million people die every year because of cardiovascular disease which counts for 31\% of all the deaths  occurring  each  year. Monitoring HR is  very  important  because  it can  provide  crucial  information  about  the  health  of  the  person  and can help in diagnosing heart problem at early stages.  Also, these vital signs can provide information  about  different  sleep  stages  as  they change in different stages. Owing to the importance of vital signs during sleep, many works have been conducted in recent days to monitor the vital signs. In this section, we review some of the latest works done for vital signs monitoring using the ubiquitous approach.

\subsubsection{Wearable-Based Approaches for Vital Signs Monitoring}
This section presents the wearable-based works for vital signs monitoring during the sleep.
Sharma \& Kan \cite{sharma2018sleep} proposed an RF-based technique for vital signs monitoring using a single passive RFID tag attached to the shirt of the user at the chest area. This work uses the near-field coherent sensing (NCS) concept for extracting the HR and RR without requiring the skin contact. Different filters are applied to extract the HR and RR from the phase values of the reflected signals. Melici et al. \cite{milici2018wireless} presented a magnetometer-based system which can monitor the breathing rate by measuring the variations in magnetic field. The magnetometer is packaged in a wearable device along with an RFduino and is worn around the body using an elastic chest belt. After applying different filters, a peak detection algorithm is used to calculate the respiration rate. The performance of the proposed system is compared with a temperature airflow sensor and the results indicate that the proposed system can monitor the RR with reasonable accuracy. 
RF-ECG  \cite{wang2018rf} is another technique which uses passive RFID tags attached to the shirt (at chest area) of the user for estimating the heart rate variability. The reflected signal from the passive tags is cleaned by removing the affect of breathing and the signals from multiple tags are fused together to strengthen the heartbeat signal. After applying other filtering techniques such as wavelet-based denoising, a PCA-based template mechanism is applied to estimate the inter-beat intervals. 
Hussain et al. \cite{hussain2019cost} proposed an RFID-based technique for sleep monitoring. Two passive RFID tags are attached to user's shirt and the reflected signals from both the tags are captured by the reader. After removing the ambient noise, the reflected signals from both tags are fused together to strengthen the breathing signal. This work uses automatic multiscale-based peak detection
(AMPD) algorithm to estimate the breathing rate. The results of the experiments show that the proposed technique can achieve good accuracy for single as well as multiple users. TagBreath \cite{wang2020tagbreathe} is another work which 
uses passive RFID tags for breath rate monitoring. TagBreath uses three tags attached to the user's shirt for capturing the chest motion resulting from breathing. Unlike \cite{hussain2019cost} which uses the RSSI values, TagBreath uses the phase values from the reflected signal and fuses the signals from all three tags. After applying multiple filters in both frequency and time domain, the breath rate is estimated by calculating the zero crossing in the selected window.

\subsubsection{Non-Wearable-Based Approaches for Vital Signs Monitoring}
In this section, we present the non-wearable-based studies for vital signs monitoring during the sleep.
SleepSense \cite{lin2017sleepsense} is a non-contact sleep monitoring system which can monitor the breathing rate along with other factors such as bed exit and on-bed movements. This work uses a Doppler radar placed above the bed which emits a single-tone carrier signal and uses the phase shifts in the reflected signal as a results of the chest movements due to respiration, to estimate the breathing rate. The phase shift is proportional to the movement displacement which can be calculated by demodulating the phase using the an extended differentiate and cross multiple (DACM) algorithm. Different statistic, time, and frequency domain feature are extracted and fed into a decision tree classifier which detect one of the three classes; breathing section, bed exit, or on-bed movement. An adaptive-threshold based peak detection algorithm is used to calculate the breathing rate in the breathing section. 
Yang et al. \cite{yang2018multi} proposed a technique for respiration monitoring using COTS WiFi devices. This work can estimate the breathing rates for multiple persons simultaneously. The key focus of this work is the optimal deployment of the WiFi transceivers such that each person is at good location of only one transceiver and bad location for the other transceiver resulting in the simultaneous monitoring of multiple persons each by a separate transceiver. 

The work conducted by Sadek \& Mohktari \cite{sadek2018nonintrusive} uses fiber optic embedded in a sensor mat to monitor different sleeping factors including RR and HR. The proposed system uses the mean and standard deviation to identify the sleep movement and empty bed states. In the sleep state, a Chebyshev band-pass filter is used to extract RR and HR signals. The HR is calculated using the Maximal overlap discrete wavelet transform while the RR is calculated using the peak detection algorithm after smoothing the signal using different filters. Helena \cite{9127369} is a recent work which uses geophone (seismic sensor) integrated in the bed frame to monitor sleep activities including the RR and HR. Helena first determines the bed status and if the user is on the bed and there is no movement, the RR and HR are then calculated using an envelop-based estimation method. After obtaining the envelop, HR can be calculated using the peak detection algorithm. Since the seismometer is insensitive to lower frequency measurements (in case of respiration), RR cannot be directly estimated from the seismic data but instead the detected HR peaks are used to generate the respiration modulation signal with the help of extrapolation technique. Then the RR can be estimated by counting the peaks from the obtained envelop of the modulated signal. The authors have done extensive experiments to evaluate the performance of Helena. Experiments have been conducted in lab environments as well as in the wild settings and the results show that Helena can monitor the vitals signs with high accuracy and in real time. Turppa et al. \cite{turppa2020vital} uses the frequency modulated continuous wave (FMCW) radar for vital sign monitoring during sleep. This work uses Fast Fourier Transform (FFT)-based cepstral analysis method for extracting the HR. For the RR, an auto-correlation function (ACF) is applied on the phase signal of a selected range. The proposed system is evaluated in different scenarios resembling real-world situations including normal and abnormal cases.

\subsubsection*{\textbf{Summary}}
A summary of the papers reviewed for vital signs monitoring is given in Table~\ref{tab:table8} and Table~\ref{tab:table9}. This category of sleep monitoring is dominated by wearable-based approach. Wearable solutions have high accuracy with low cost. RF technology is widely used in the wearable solutions in which passive RFID tags are attached to chest or shirt of the users. Non-wearable solutions are also using the RF technology in the form of radars and radio devices installed in the room to capture the sleep data. Unlike the other categories of sleep monitoring which mostly use the machine learning, signal processing is used for vital signs estimation. One of the main focus of the presented works is the use of novel methods for removing the noise from the captured data because the vital signs can easily be suppressed by the environment noise. Various filtering techniques have been presented to clean the signal and extract the RR and HR of the sleeping person. One concern here is the amount of data collected for the evaluation purposes. Most of the experiments presented in these works are simulating the sleep and not the actual sleep, i.e., the users are resting/sitting to simulate the sleep. There is a need of long experiments which involve the overnight sleep of the user to reflect the real world scenarios.

\begin{table}[!bt]
\centering
\begin{adjustbox}{max width= \linewidth}
\begin{tabular}{|p{1.5 cm}|p{1.5 cm}|p{2.5 cm}|p{2.5cm}|p{3 cm}|p {1.5 cm}|l|p{1 cm}|p{2 cm}|p{1.5 cm}|p{2.5 cm}|}
\hline
\textbf{Work}                                   & \textbf{Vital Signs} & \textbf{Sensor-Placement} & \textbf{Device Used}      & \textbf{Methods Used}                                         & \textbf{Accuracy} & \textbf{Obtrusiveness} & \textbf{Cost} & \textbf{Parameter Observed} & \textbf{\# Subjects} & \textbf{Validation Method}     \\ 
\hline
\cite{sharma2018sleep} 2018    & HR,RR                & RFID-chest                & NA                        & Signal processing                                             & 97.58\%           & Low                    & Low           & RSSI                        & 1                    & Observation                    \\ 
\hline
\cite{milici2018wireless} 2018 & RR                   & Magnetometer-chest        & Self-made                 & Signal processing                                             & 85\%              & Medium                 & Low           & Magnetic vectors            & 1                    & Air-flow sensor                \\ 
\hline
\cite{wang2018rf} 2018         & HRV                  & RFID tags-shirt           & ImpinJ tags and reader    & Signal processing, DWT, fusion, PCA-based template estimation & 93\%              & Low                    & Low           & Phase values                & 15                   & Heal Force PC-80A ECG Monitor  \\ 
\hline
\cite{hussain2019cost} 2019    & RR                   & RFID tags-shirt           & Alien tags and reader     & Signal processing                                             & 95\%              & Low                    & Low           & RSSI                        & 4                    & Observation                    \\ 
\hline
\cite{wang2020tagbreathe} 2020   & RR                   & RFID tags-shirt           & Impinj reader, Alien tags & Signal processing                                             & 95\%              & Low                    & Low           & Phase values                & 4                    & Metronome mobile application   \\
\hline
\end{tabular}
\end{adjustbox}
\caption{Comparison of the wearable-based solutions for vital signs monitoring}
\label{tab:table8}
\vspace{-5mm}
\end{table}

\begin{table}[!bt]
\centering
\begin{adjustbox}{max width= \linewidth}
\begin{tabular}{|p{1.5 cm}|p{1.5 cm}|p{2.5 cm}|p{1.5cm}|p{3.5 cm}|p {2.5 cm}|l|p{1.5 cm}|p{2.5 cm}|p{1.5 cm}|p{2.5 cm}|}
\hline
\textbf{Work}                                      & \textbf{Vital Signs} & \textbf{Sensor-Placement}  & \textbf{Device Used}            & \textbf{Methods Used}                                                                                                 & \textbf{Accuracy}                      & \textbf{Obtrusiveness} & \textbf{Cost} & \textbf{Parameter Observed}                        & \textbf{\# Subjects} & \textbf{Validation Method}                                        \\ 
\hline
\cite{lin2017sleepsense} 2017     & RR                   & Doppler Radar- above bed   & Self-made                       & Decision tree                                                                                                         & 93.45\%                                & Low                    & Low           & Phase shift in the reflected signal                & 3                    & Nasal airflow sensor                                              \\ 
\hline
\cite{yang2018multi} 2018         & RR                   & WiFi-bedside               & 5300 NIC, WR841N router & Gaussian movement model, Fresnel zone, hampel and wavelet filters, peak detection & 1 bpm(MAE)                             & Low                    & Low           & Changes caused in received signals by chest motion & 5                    & Chest
strap                                                       \\ 
\hline
\cite{sadek2018nonintrusive} 2018 & RR,HR                & Optical fiber-sleeping mat & Self-made                       & Signal processing, maximal overlap
discrete wavelet transform, peak detection                                         & NA                                     & Low                    & Low           & pressure exerted by body                           & 3                    & Self-reported survey                                              \\ 
\hline
\cite{9127369} 2020               & RR,HR                & Seismic
sensor-bedframe    & Self-made                       & Envelope based HR/RR estimation, peak detection                                                                        & 1.26 bpm,0.52 rpm- 2.41 bpm,NA (MAE),. & Low                    & Low           & Pressure exerted by body                           & 10+25+2              & pulse
oximeter( Zorvo), metronome, Apple watch                      \\ 
\hline
\cite{turppa2020vital} 2020       & RR,HR                & FMCW radar-above the bed   & Self-made                       & Signal processing, peak etection, haming window, autocorrelation function, Hann window                                & 3.6\% for HR,9.1\% for RR (MAE)        & Low                    & Low           & Changes caused in received signals by chest motion & 10                   & PSG by Embla Titanium and VTT's BCG installed under the bedsheet  \\
\hline
\end{tabular}
\end{adjustbox}
\caption{Comparison of the non-wearable-based solutions for vital signs monitoring}
\label{tab:table9}
\end{table}

\subsection{Sleep Disorders}

Sleep disorders are different diseases which are attributed to lack of quality sleep and one of the main reasons of inadequate sleep problem. Sleep disorders also contribute to other diseases. Some of the most common symptoms of sleep disorders are irritability, depression or anxiety, tiredness, and daytime sleepiness \cite{senaratna2017prevalence}. 
According to the International Classification of Sleep Disorders (ICSD-3), there are seven types of sleep disorders which are sleep-related breathing disorders, insomnia, sleep-related movement disorders, circadian rhythm sleep-wakefulness
disorders, central hypersomnolence disorders, parasomina and other disorders \cite{pan2020current}. These disorders can be cured with proper treatment but the diagnosis of these disorders is a challenging task. Recently, many in-home techniques have been proposed which can provide comparable results with clinical methods. In this section, we review some of the latest works conducted for detection of sleep disorders using the unobtrusive approach.

\subsubsection{Wearable-based Approaches for Sleep Disorder Detection}
In this section, we review the wearable-based techniques used for sleep disorders detection.
Kye et al. \cite{kye2017detecting} presented a wearable band-based system to detect periodic limb movements (PLMS). This band consists of inertial sensors (accerelometer and gyroscope) and is worn on the dorsum of the foot. This work proposes the motion synchronized windowing (MSW) technique in the feature extraction phase in which the data stream is segmented into intervals where the motion occurs. 
Different classifier are trained and tested but KNN achieves the highest accuracy for detection of PLMS in the performed experiments. AutoTag \cite{yang2018autotag} uses the passive RFID tags attached to the chest area of the user for detection of sleep apnea. This work proposes a novel technique for mitigating the frequency hopping offset of RFID system. Unlike the other approaches, AutoTag uses an unsupervised recurrent autoencoder method for apnea detection
and 
treats the breathing signal as a time sequence in a time window and uses an LSTM for encoding the breathing signal. The output of the LSTM is used to estimate the mean and variance vector which in turn are used for computing the latent vector. Another LSTM is used for decoding the mean and variance vectors to get the reconstructed breathing signal. Kullback Leibler (KL) divergence method is used to detect apnea by comparing the difference between original and reconstructed signals. 

Fang et al. \cite{fang2018novel} proposed an apnea detection system based on the characteristic moment waveform (CMW) method. This work uses a microphone attached near the nose of the user to capture the acoustic signals when the user breaths. In the prepossessing stage, the amplitude contrast of the original signal is reduced by using different signal processing techniques and then CMW is used for RR estimation based on time characteristic waveform (TCW). Jeon \& Kang \cite{jeon2019wearable} developed a wearable sleepcare kit which can detect sleep apnea in real-time. This kit consists of two parts: a sensing device called Bio-Cradle and a control device called personal activity assisting
\& reminding (PAAR) band. The Bio-cradle consists of PPG and accelerometer sensors and a nasal tube for collecting physiological data while the PAAR works as a scheduler.  Sleepcare kit is a plug and play system which doesn't require any prior training data and can monitor the sleep in real time using a mobile application. This system uses the breathing and SpO2 level to detect apnea and hypopnea. The proposed system can also identify the level of apnea and can alert the caregivers in real-time using the mobile application. Bobovych et al. \cite{bobovych2020resteaze} also developed a wearable device which can be worn on ankle position and can capture two types of leg movements: dorsiflexion (which are correlated with restless leg syndrome (RLS) and Attention Deficit Hyper Activity Disorder (ADHD)) and complex leg movements (which are correlated with sleep fragmentation and brief arousals). Since the proposed system was supposed to be resource constraint (wearable), a single feature (standard deviation) is used to train an SVM classifier. The main focus of this work is not the processing part but rather the power usage analysis  of the hardware and software.

\subsubsection{Non-Wearable-based Approaches for Sleep Disorder Detection}

This section presents the non-wearable-based solutions for sleep disorders detection.
Nandakumar et al. \cite{nandakumar2015contactless} presented a contact-less solution for apnea detection using the built-in speakers in the smartphones. The speaker in the phone is used as a sonar by emitting a frequency-modulated sound signal and the reflection is captured and analyzed to extract the breathing information of the user sleeping near by. A modified peak detection algorithm is used to detect different sleeping events such as central apnea, obstructive apnea and hypopnea based on the breathing pattern. Waltisberg et al. \cite{waltisberg2017detecting} presented a framework for the detection of sleep apnea and PLMS using the in-bed sensor system which consists of strain gauges to detect the pressure changes caused by the breathing motion. This work explores both the measurement fusion and the decision fusion approaches. PCA and a two-layered feature extraction approach is used while minimum redundancy
maximum relevance criterion (mRMR) filter is used for feature selection. An NB classifier is used to classify the sleeping event.
 The results show that both measurement and decision fusion approaches give same results and are better as compared to a rule-based approach using the standard deviation. 
 
 EZ-Sleep \cite{hsu2017zero} uses an ambient radio device installed in the bedroom to monitor insomnia and other sleep activities. The radio device emits RF signals which are reflected from the human body and the environment. EZ-Sleep analyzes these reflected signal to extract the sleep-related information such as bed location, bed-entry, bed-exit, and other insomnia assessment parameters. The bed is located first using 
 a 2D histogram and image segmentation technique. Once the bed is located, the next step is to identify whether the user is lying on the bed or not. EZ-Sleep achieves this by using Hidden Markov Model (HMM). Before estimating the sleep parameters (vital signs etc.), EZ-Sleep first needs to detect when the user fall asleep (sleep vs awake classification). This is done by using CNN with 14-layer residual network model and to further increase the accuracy for detecting onset sleep, Gradient Boosting Regressor (GBR) is build on the top of CNN. Results show that EZ-Sleep can detect different sleep activities with high accuracy. 
 Swangarom et al. \cite{swangarom2018proposal} proposed a contact-less system for insomnia detection using the fabric-pressure sensors embedded in the sleeping mat. The sensors can capture the body movements resulting from breathing and changing positions. The Athens insomnia scale (AIS) questionnaire is used for validation of the results. An SVM with polynomial kernel function is used for classifying the subjects into two classes; healthy and suffering from insomnia. Sadek et al. \cite{sadek2020new} proposed another non-invasive solution for apnea detection using the fiber optic sensors embedded in a mat and put under the mattress. This work uses the BCG principle and captures the cardiac movements (HR) and chest movements (RR). After applying multiple signal processing techniques to clean and strengthen the signal, an adaptive
histogram-based thresholding method is used for apnea detection.

\begin{table}
\centering
\begin{adjustbox}{max width= \linewidth}
\begin{tabular}{|p{1.5 cm}|p{1.7 cm}|p{2.5 cm}|p{2.5cm}|p{2.5 cm}|p {1.5 cm}|l|p{1.5 cm}|p{2.5 cm}|p{1.5 cm}|p{2 cm}|}
\hline
\textbf{Work}                                     & \textbf{Disorders} & \textbf{Sensor-Placement}               & \textbf{Device Used}            & \textbf{Methods Used}  & \textbf{Accuracy} & \textbf{Obtrusiveness} & \textbf{Cost} & \textbf{Parameter Observed}                         & \textbf{\# Subjects} & \textbf{Validation Method}      \\ 
\hline
\cite{kye2017detecting} 2017     & PLMS               & Accelerometer, gyroscope-foot dorsum    & MetaMotionC                     & KNN                    & 96.92\%           & Low                    & Low           & Foot movements                                      & 13                   & Observation                     \\ 
\hline
\cite{yang2018autotag} 2018      & Apnea              & RFID tags-chest                         & Impinj R420, reader, Alien tags & LSTM, KL divergance    & 92\%              & Low                    & Low           & Breathing from phase values of the reflected signal & 4                    & NEULOG Respiration Belt Sensor  \\ 
\hline
\cite{fang2018novel} 2018        & OSA                & Microphone-near nose                    & Self-made                       & Signal processing      & 97.00\%           & Medium                 & Medium        & Acoustic signal from breathing                      & 5                    & Manually                        \\ 
\hline
\cite{jeon2019wearable} 2019     & Apnea, hypopnea     & Accelerometer, PPG-wrist, nasal airflow & Self-made                       & Signal processing, CMW & NA                & Medium                 & Medium        & SpO2, HR, RR                                        & 15                   & PSG                             \\ 
\hline
\cite{bobovych2020resteaze} 2020 & RLS                & Capacitive and inertial sensor-ankle    & Self-made                       & SVM with linear kernel & 90\%              & Low                    & Medium        & Leg movements                                       & 12                   & PSG, IR camera                  \\
\hline
\end{tabular}
\end{adjustbox}
\caption{Comparison of the wearable-based solutions for disorders detection}
\label{tab:table10}
\end{table}

\subsubsection*{\textbf{Summary}}
Table~\ref{tab:table10} and Table~\ref{tab:table11} provide a summary of the work presented for sleep disorders detection. As can be seen from the tables, significant works have been conducted in sleep disorder detection especially in apnea detection. Wearable-based solutions have overall better accuracy. Both machine learning and signal processing are used to detect the disorders. Getting the ground truth for movement related disorders is one of the concerns in this category. Again, most of the experiments are simulating the sleep and apnea episodes rather than involving actual patients who are suffering from these disorders.  

\begin{table}
\centering
\begin{adjustbox}{max width= \linewidth}
\begin{tabular}{|p{1 cm}|p{2.5 cm}|p{2.5 cm}|p{2.5cm}|p{3 cm}|p {1.5 cm}|l|p{1 cm}|p{2 cm}|p{1.5 cm}|p{2 cm}|}
\hline
\textbf{Work}                                          & \textbf{Disorders}                             & \textbf{Sensor-Placement}    & \textbf{Device Used}                  & \textbf{Methods Used}                               & \textbf{Accuracy} & \textbf{Obtrusiveness} & \textbf{Cost} & \textbf{Parameter Observed} & \textbf{\# Subjects} & \textbf{Validation Method}  \\ 
\hline
\cite{nandakumar2015contactless} 2015 & Central apnea, obstructive apnea, and hypopnea & Sonar-bedside                & Four different smarphones             & Signal processing, FMCW, peak detection             & 97\%              & Low                    & Low           & Chest and abdomen motions   & 37                   & PSG + Video camera          \\ 
\hline
\cite{waltisberg2017detecting} 2017   & Apnea, hypopnea, PLM                           & Strain guage sensors-bed     & Mobility monitor by compliant concept & Fusion architecture, mRMR filter, NB                & 72\%              & Low                    & High          & Pressure changes            & 9                    & PSG                         \\ 
\hline
\cite{hsu2017zero} 2017               & Insomnia                                       & Radio-room                   & Self-made                             & Signal processing, FMCW, image processing, HMM, CNN & NA                & Low                    & Low           & Reflected signal            & 10                   & SleepProfiler               \\ 
\hline
\cite{swangarom2018proposal} 2018     & Insomnia                                       & Pressure sensors-bedsheet    & Self-made                             & SVM with polynomial kernel                          & 77.78\%           & Low                    & Low           & Body pressure               & 9                    & AIS questionnaire           \\ 
\hline
\cite{sadek2020new} 2020              & Apnea                                          & Fiber optic sensors-bedsheet & Self-made                             & Adaptive
histogram-based thresholding               & 49.96\%           & Low                    & Low           & Cardiac and chest movements & 10                   & PSG                         \\
\hline
\end{tabular}
\end{adjustbox}
\caption{Comparison of the wearable-based solutions for disorders detection}
\label{tab:table11}
\end{table}

\section{Open Issues and Future Directions} \label{sec:5}
Although significant work has been conducted in sleep monitoring using the unobtrusive and in-home solutions, there are still some open issues in this area which need to be addressed. In this section, we discuss some of those open issues and provide future research directions.

\subsubsection*{Trade-off between Accuracy and Comfort Level}
Accuracy is always an important factor for any solutions. It is more important in areas like sleep monitoring which consists of vital signs monitoring and sleep disorder detection. From the related literature we have concluded that there is a trade off between the accuracy and the comfort level (unobtrusiveness) of the proposed solutions. Solutions which are more invasive in nature, provide better accuracy for all the four categories of sleep monitoring. On the other hand, solutions which are non-invasive and are easy to use, their accuracy is low in comparison especially in sleep stages classification and disorder detection. These in-home approaches can be used for general sleep monitoring but more work is required to bring the accuracy level to medical grade monitoring. There is a potential for further research to have solutions which are non-invasive and have the same level of accuracy as the clinical solutions.

\subsubsection*{Security and Privacy} 
In sleep monitoring, different sensors (wearable and non-wearable) collect the physiological data of the sleeping person. This data is then sent to either an edge device 
or 
to a central module (could be cloud) for further processing (recognition). In some cases, the processing entity is installed along with the sensors (in form of microcomputers, aurdinos, etc.) and the results are transferred via the network. From the related literature, we can see that the focus is mainly on the accuracy and ease of use. Very little (and mostly none) efforts are made for the security and privacy of the collected data. Most of the data is medical data and should not be shared with unauthorized person but no specific measures are taken to protect the data from unauthorized access. Since the areas like telemedicine and remote monitoring are gaining popularity in the healthcare, there is a need for adapting specific measure to protect the data. Hathaliya \& Sudeep \cite{hathaliya2020exhaustive} studied the security and privacy issues in health care and identified some of the key challenges.

\subsubsection*{Limited Data}
Data is a main aspect of sleep monitoring. Publicly available data sets are very important for evaluating and comparing the performance of any newly proposed solution. From the literature, we found that there are some data sets publicly available but most of them are of PSG data collected in a clinical environment. These data sets can help in the performance evaluation of solutions which use the same technique i.e., PSG or ECG etc., but nowadays many solutions are unobtrusive and use wearable and now-wearable devices for monitoring physiological signals other than PSG. These solutions are implemented in home environments and use devices such as radio, RFID, accelerometer, magnetometer, pressure sensors, and WiFi. There is no data sets publicly available for these type of approaches except \cite{pouyan2017pressure} and \cite{hillyard2018experience}. None of the related works covered in Section~\ref{sec:4} except \cite{zhao2017learning} have released their data set. Since the use of intrusive technique is increasing for sleep monitoring, there is a need for benchmark data sets to evaluate the performance of new solutions. The problem of limited data can be solved by user contributions i.e., crowdsourcing the sleep data from their in-home devices. Also, the authors can be encouraged to share their data sets for validation purpose. Hence, large data sets can be created which are more balanced and diverse to reflect the real-world scenarios. 

\subsubsection*{Need for Sleep-Specific Devices}
From the related literature, it can be seen that many devices which are used for sleep monitoring, are the commercially available general purpose devices mainly used for activity tracking. Sleep monitoring is often considered as an additional feature of these devices. In some cases, the researchers have customized these devices to use it for sleep monitoring. Acceptance by the users is one of the main aspect of wearable devices. A study conducted by Shin et al. \cite{shin2019wearable} reported that it is unreasonable to design an all-purpose wearable activity tracker. Although some sleep-specific devices exist in the market, there is a need of more research and development to produce specific sleep-monitoring devices that are cheap and easy to use for both short-term and long-term monitoring. 

\subsubsection*{Use of Machine Learning}
After collecting the physiological signals, two approaches can be used to infer the sleep related information. These approaches are signal processing and machine learning. We have found in the literature review, that the use of machine learning is increasing for sleep monitoring. Many studies have used machine learning and have explored the feature engineering to get good results. One of the advantages of using machine learning is the high accuracy for sleep monitoring especially for detecting disorders, sleep stages, and postures. However, machine learning requires more resources which is a problem for wearable devices given their small sizes and battery requirements. Some solutions have tried to tackle this problem by shifting the training phase offline \cite{jeng2021wrist} while other have used light weight techniques to ease the burden on the wearable devices.

\subsubsection*{Long-Term vs Short-Term Monitoring}
The related literature shows that unobtrusive approaches (both wearable and non-wearable) are evaluated only for the short-term monitoring (mostly a single night or a few hours). However, it would be interesting to see the performance for a long-term monitoring. Despite the low accuracy of unobtrusive approaches as compared to clinical solutions, an advantage of the unobtrusive solutions is their ease of use and low cost for long-term deployment. These solutions can be explored for long-term in-home monitoring and can help in the early diagnosis of many health relate problems such as epilepsy \cite{lucey2015new}, depression \cite{fan2019long} and other respiratory problems \cite{bachour2016oral}.

\section{Conclusion} \label{sec:6}
Sleep monitoring has been an important approach to understand sleep behaviors and improve sleep quality. With the recent advancements in sensor technologies, unobtrusive sleep monitoring using wearable or non-wearable sensors has become an active research area.  
In this paper, we have presented a comprehensive review of the latest research efforts in the area of sleep monitoring. We divide the related literature into four categories: sleep stages classification, sleep postures recognition, sleep disorders detection, and vital signs monitoring. Unlike the previous surveys which focus on some of the categories in sleep monitoring, we cover all the four categories. The main focus of this survey is the unobtrusive techniques used for in-home sleep monitoring. We review the latest work in all the four categories of sleep monitoring and provide a comparison based on 10 key factors. We also identify and discuss some open research issues and provide future research directions
 to stimulate further research in this important topic.

\bibliographystyle{ACM-Reference-Format}
\bibliography{sample-base}

\appendix

\end{document}